\documentclass[aps,pre,showpacs,amsmath,amssymb,reprint]{revtex4-1}

\usepackage{bm}
\usepackage{graphicx}

\begin{document}

\title{Numerical verification of random phase-and-amplitude formalism of weak turbulence}

\author{Mitsuhiro Tanaka}
\email{tanaka@gifu-u.ac.jp}
\affiliation{Faculty of Engineering, Gifu University, Gifu 501-1193 Japan}

\author{Naoto Yokoyama}
\email{yokoyama@kuaero.kyoto-u.ac.jp}
\affiliation{Department of Aeronautics and Astronautics, Kyoto University, Kyoto 615-8540, Japan}

\date{\today}

\begin{abstract}
The Random Phase and Amplitude Formalism (RPA)
has significantly extended the scope of weak turbulence studies.
Because RPA does not assume any proximity to the Gaussianity in the wavenumber space,
it can predict, for example, how the fluctuation of the complex
amplitude of each wave mode grows through nonlinear interactions with other modes,
and how it approaches the Gaussianity.
Thus, RPA has a great potential capability,
but its validity has been assessed neither numerically nor experimentally.
We compare the theoretical predictions given by RPA with the results of direct numerical simulation (DNS)
for a three-wave Hamiltonian system, thereby assess the validity of RPA.
The predictions of RPA agree quite well with the results of DNS
in all the aspects of statistical characteristics of mode amplitudes studied here.
\end{abstract}


\maketitle

\section{Introduction}

In weak turbulence,
wave trains which have different directions of the propagation, wavelengths, and frequencies
weakly interact with each other
owing to the nonlinearity in the governing equation and/or the boundary conditions.
Many papers on weak turbulence have been devoted to
the derivation of the kinetic equation~\cite{hasselmann,zak_filo1967,Lvov2003333}, which governs the statistical evolution of the wave action spectrum,
the physical and mathematical properties of the kinetic equation and resulting spectra~\cite{Connaughton200386,kartashova2010nonlinear,newell01,nls_intermittent_cycle,ISI:000287046400003},
the statistically steady non-equilibrium spectra when the external forces and the dissipation balances~\cite{pushkarev_1999,during2006weak},
and so on.
Thus, researches on weak turbulence have been conventionally focused on the wave action spectrum,
which is the ensemble average of the squared norm of the complex amplitude of each mode~\cite{zak_book}.
On the other hand,
the spectra which are observed in the direct numerical simulations (DNS) and field observations have large fluctuations.
In DNS, the amplitude of each wave mode is often determined by a prescribed spectrum
and initially have no fluctuations at all.
Even in such DNS, the amplitude fluctuations spontaneously grow as time elapses.
The generation mechanism of the fluctuations, the time scales of their growths,
and the possibility of the approach to the Gaussianity
had little been studied.

The Random Phase and Amplitude Formalism (RPA) which has recently been developed
has changed the situation, and drastically extended the scope
of weak turbulence research~\cite{lvov2004nsl,nazarenkobook,Choi2005361,Eyink20121487}.
Similar to the conventional weak turbulence theory,
RPA is the statistical theory for the complex amplitude $a_{\bm{k}}$.
Let $a_{\bm{k}}$ be expressed as $a_{\bm{k}}=|a_{\bm{k}}| \psi_{\bm{k}}$
 with the positive amplitude $|a_{\bm{k}}|$ and the phase factor $\psi_{\bm{k}}= e^{i\phi_{\bm{k}}}$.
In the Random Phase approximation (RP) in the strict sense, it is assumed that $\psi_{\bm{k}}$ for all $\bm{k}$ are
independent random variables and are uniformly distributed over the unit circle in the complex plane.
In the Random Phase and Amplitude Formalism it is assumed, in addition to the assumptions of RP,
that the amplitude $|a_{\bm{k}}|$ also are mutually independent random variables for all $\bm{k}$.
It is important to note that in RPA $|a_{\bm{k}}|$ is allowed to have any distribution
and can be far from the Gaussianity.
Although it is only recently that RPA is defined unambiguously,
it has implicitly been used for a rather long time under the name of \lq\lq the random phase approximation",
often without being realized of it.
For example, Zakharov et al. (\cite{zak_book}p.65) calls their approximation \lq\lq the random phase approximation",
but they implicitly assume the statistical independence of the amplitude $|a_{\bm{k}}|$ for different $\bm{k}$.

The equations which describe the time evolution of the moments
of $|a_{\bm{k}}|^2$ for arbitrary orders
as well as the probability density function (PDF) of $|a_{\bm{k}}|^2$
have been derived.
Thus, RPA is expected to describe the statistical characteristics of the amplitude fluctuations,
which have little been studied.
Although RPA seems to have a great potential capability,
its validity has been assessed neither numerically nor experimentally.
In this study, we perform a series of large-scale DNS
for a model Hamiltonian system which allows three-wave resonance,
and compare the results with the theoretical predictions of RPA.
Here we confine our attention to the single-mode statistics of the amplitude fluctuations.
In every aspect studied here, we have obtained good quantitative agreement between RPA and DNS.

Benney and Newell~\cite{benney1969random} investigated the $n$th order cumulant $R^{(n)}$ of a wave turbulence field in the  physical space,
and derived an equation which governs the temporal evolution of the Fourier transform of $R^{(n)}$.
They assumed that the wave field is weak nonlinear and that the medium is dispersive, but did not assume that the wave
field is close to a Gaussian state in the physical space.
Prior to \cite{benney1969random}, Benney and Saffman~\cite{Benney25011966} derived the kinetic equation for the action spectrum based on the same assumptions.
In the present paper, we will compare our numerical results with the predictions of RPA only, and will not try any comparison with
the predictions of \cite{benney1969random} mainly from the following two reasons.
Firstly, \cite{benney1969random} is described in terms of the Fourier transform of a real-valued physical variable, and it is not
expressed in terms of the complex amplitude of wave modes which we want to handle.
Secondly, \cite{benney1969random} does not introduce the discretization of the $\bm{k}$ space
and the Fourier transform is a generalized function.  On the other hand, RPA treats the discretized $\bm{k}$ space from the outset,
hence the comparison with DNS is straightforward and much easier than the case of \cite{benney1969random}.
There is also an essential difference between RPA and the analysis in \cite{benney1969random},\cite{Benney25011966} with regard to the
speed of decay of the correlations in the physical space as functions of the separation of points. (See for example \cite{lvov2004nsl}.)

This paper is organized as follows.
The numerical scheme is presented in Sec.~\ref{sec:numericalscheme}.
In Sec.~\ref{sec:numerical result},
numerical results of the evolution of the amplitude fluctuations and the approach to the Gaussianity
are reported.
The discussion and the summary are given in Sec.~\ref{sec:concludingremark}.

\section{Numerical Scheme}
\label{sec:numericalscheme}
\subsection{Numerical Model}
In this study we employ the following three-wave Hamiltonian system:
\begin{subequations}
\begin{align}
\mathcal{H} &= \mathcal{H}_2+\mathcal{H}_3, 
\\
\mathcal{H}_2 &= \int \omega(\bm{k})|a(\bm{k})|^2 d\bm{k}, 
\label{eqn:H2}
\\
\mathcal{H}_3 &= \frac{1}{2}
\int
\left(
 V(\bm{k}, \bm{k}_1, \bm{k}_2) a^{\ast}(\bm{k}) a(\bm{k}_1) a(\bm{k}_2)+\mathrm{c.c.}\right)
\nonumber\\
&
\qquad\qquad
\times
\delta(\bm{k} - \bm{k}_1 - \bm{k}_2) d\bm{k}_{123}
.
\label{eqn:H3}
\end{align}
\label{eq:hamiltonian}
\end{subequations}
\begin{widetext}
\begin{subequations}
\begin{align}
\frac{d a(\bm{k})}{dt}
&= -i\frac{\delta \mathcal{H}}{\delta a^{\ast}(\bm{k})} \nonumber\\
&= -i\omega(\bm{k}) a(\bm{k}) 
\nonumber\\
& \quad
-\frac{i}{2} \int V(\bm{k}, \bm{k}_1, \bm{k}_2)
 a(\bm{k}_1) a(\bm{k}_2) \delta(\bm{k} - \bm{k}_1 - \bm{k}_2) d\bm{k}_{12}
 -i \int V^{\ast}(\bm{k}_1, \bm{k}, \bm{k}_2)
 a(\bm{k}_1) a^{\ast}(\bm{k}_2) \delta(\bm{k}_1 - \bm{k} - \bm{k}_2) d\bm{k}_{12},
\label{eqn:continuous 3wave equation}
\\
&
\omega(\bm{k}) = k^{\alpha},
 \quad
 V(\bm{k}, \bm{k}_1, \bm{k}_2) = (k k_1 k_2)^{\beta},
 \quad 
 \alpha = 3/2,
 \quad
 \beta = 1/4.
\label{eqn:model_2}
\end{align}
\label{eq:dynamic}
\end{subequations}
\end{widetext}
Here, $\bm{k}$ is a two-dimensional wavenumber vector,
$a^{\ast}$ expresses the complex conjugate of $a$,
and $\mathrm{c.c.}$ also expresses the complex conjugate of the preceding term. 
The linear frequency and the complex amplitude of the mode of the wavenumber $\bm{k}$
are respectively expressed by $\omega(\bm{k})$ and $a(\bm{k})$.
The short-hand notations $k=|\bm{k}|$ and $d\bm{k}_{12}=d\bm{k}_1 d\bm{k}_2$
are used.
When we derive the dynamic equations for the complex amplitude in weak turbulence systems,
we often obtain the dynamic equation like Eq.~(\ref{eqn:continuous 3wave equation})
when the three-wave resonant interactions are allowed like in the surface capillary waves~\cite{zak_book}.
In fact,
the difference between the dynamic equation for surface capillary waves and that for our model
appears only in the interaction kernel $V(\bm{k}, \bm{k}_1, \bm{k}_2)$.
For our objective to generally compare DNS with RPA,
we select the simple interaction kernel that allows the evaluation of the convolution in the nonlinear terms
to be performed fast by the Fast Fourier Transforms (FFT).

In numerical studies of weak turbulence, one sometimes adds artificial energy input and/or output
to the conservative system (\ref{eqn:continuous 3wave equation}).
If we were to investigate weak turbulence characteristics in a statistically steady state, such as
the Kolmogorov--Zakharov spectrum, it would be necessary to add such non-conservative effects.
On the other hand, the purpose of the present study is to assess the validity of the prediction
of RPA on the temporal evolution of various statistics of $a_{\bm{k}}$.
Therefore, we need statistical unsteadiness of the wave field, and in this respect the conservative system
without input or output is in accord to our purpose as it stands.

\subsection{Correspondence between Continuous System and Discrete System}
Some cautions should be exercised
when we compare the theoretical description where the wavenumbers are continuous
with numerical results where the wavenumbers are discrete.
To connect the wavenumber space and the real space,
we select the definition of the Fourier transform as follows: 
\begin{align}
f(\bm{x}) &= \frac{1}{2\pi} \!\int\! F(\bm{k}) e^{i\bm{k}\cdot\bm{x}} d\bm{k},
\quad
F(\bm{k}) = \frac{1}{2\pi} \!\int\! f(\bm{x}) e^{-i\bm{k}\cdot\bm{x}} d\bm{x},
\end{align}
where $\bm{x}$ is a two-dimensional vector in the real space.
We also select the Fourier series
which connects the rectangular domain $R = L_x \times L_y$ in the real space
under the doubly periodic boundary conditions
to the discrete wavenumber $\bm{k}$:
\begin{align}
f(\bm{x}) &= \sum_{\bm{k}\in S_{\bm{k}}} F_{\bm{k}} e^{i\bm{k}\cdot\bm{x}},
\quad
F_{\bm{k}} = \frac{1}{L_x L_y}\int_R f(\bm{x}) e^{-i\bm{k}\cdot\bm{x}} d\bm{x},
 \end{align}
where $S_{\bm{k}}$ is a set of $\bm{k}$ allowed in the discrete wavenumber space, i.e.,
\begin{equation}
 S_{\bm{k}} = \left\{\bm{k} \mid \bm{k} = (m\Delta k_x,n\Delta k_y),
	       (m,n) \in \mathbb{Z}^2 \right\}
.
\end{equation}
The grid intervals in the discrete wavenumbers $\Delta k_x$ and $\Delta k_y$
are connected to the periods in the real space $L_x$ and $L_y$ as
\begin{equation}
\Delta k_x=\frac{2\pi}{L_x}, \quad \Delta k_y=\frac{2\pi}{L_y}
.
\end{equation}

Because of the properties of the delta function
\begin{align}
\delta(\bm{k})=\frac{1}{(2\pi)^2}\int e^{i\bm{k}\cdot\bm{x}} d\bm{x},
\end{align}
and of the Kronecker's delta for $\bm{k}\in S_{\bm{k}}$
\begin{align}
\delta_{\bm{k},\bm{0}} = \frac{1}{L_x L_y}\int_R e^{i\bm{k}\cdot\bm{x}} d\bm{x}
= \frac{\Delta k_x \Delta k_y}{(2\pi)^2}\int_R e^{i\bm{k}\cdot\bm{x}} d\bm{x},
\end{align}
the correspondence
\begin{equation}
\delta(\bm{k})
\longleftrightarrow
\frac{1}{\Delta k_x \Delta k_y}\delta_{\bm{k},\bm{0}}
\label{eqn:relation between delta}
\end{equation}
is found for sufficiently small $\Delta k_x$ and $\Delta k_y$.

Because of the relation between $F(\bm{k})$ and $F_{\bm{k}}$
which generally holds
\begin{equation}
F(\bm{k}) = \sum_{\bm{k}^{\prime}\in S_{\bm{k}}} 2\pi F_{\bm{k}^{\prime}} \delta(\bm{k}-\bm{k}^{\prime})
,
\end{equation}
and the correspondence~(\ref{eqn:relation between delta}),
the following correspondence is found:
\begin{equation}
a(\bm{k}) \longleftrightarrow \frac{2\pi}{\Delta k_x \Delta k_y} a_{\bm{k}}
,
\label{eqn:relation between a}
\end{equation}
for $\bm{k} \in S_{\bm{k}}$.

The wave actions, $n(\bm{k})$ for the continuous system and $n_{\bm{k}}$ for the discrete system,
are respectively defined as
\begin{equation}
n(\bm{k})\delta(\bm{k}-\bm{k}^{\prime}) = \langle a(\bm{k}) a^{\ast}(\bm{k}^{\prime}) \rangle
\mathrm{\ and\ }
n_{\bm{k}} \delta_{\bm{k},\bm{k}^{\prime}}=\langle a_{\bm{k}}a_{\bm{k}^{\prime}}^{\ast} \rangle
,
\end{equation}
where $\langle \cdots \rangle$ represents the ensemble average.
The correspondences~(\ref{eqn:relation between delta}) and (\ref{eqn:relation between a}) give
the correspondence of the wave actions
\begin{equation}
n(\bm{k}) \longleftrightarrow \frac{(2\pi)^2}{\Delta k_x \Delta k_y}n_{\bm{k}}
.
\end{equation}

The governing equation in the discretized wavenumbers corresponding to Eq.~(\ref{eqn:continuous 3wave equation}) is written as
\begin{align}
\frac{d a_{\bm{k}}}{dt} 
=-i\omega_{\bm{k}}a_{\bm{k}} 
&
-\frac{i}{2}(2\pi)
\sum_{\bm{k}_1,\bm{k}_2}V^{\bm{k}}_{\bm{k}_1 \bm{k}_2}  a_{\bm{k}_1} a_{\bm{k}_2} \delta^{\bm{k}}_{\bm{k}_1 \bm{k}_2}
\nonumber\\
&
 -i(2\pi)\sum_{\bm{k}_1,\bm{k}_2} V^{\bm{k}_1}_{\bm{k} \bm{k}_2}  a_{\bm{k}_1} a_{\bm{k}_2}^{\ast} \delta^{\bm{k}_1}_{\bm{k} \bm{k}_2},
\label{eqn:discrete 3wave equation}
\end{align}
where $\omega_{\bm{k}} = \omega(\bm{k})$, $V^{\bm{k}}_{\bm{k}_1\bm{k}_2}=V(\bm{k},\bm{k}_1,\bm{k}_2)$,
and $\delta^{\bm{k}}_{\bm{k}_1 \bm{k}_2}$ expresses the Kronecker's delta $\delta_{\bm{k},\bm{k}_1+\bm{k}_2}$.
Note that,
in addition that the integration and the delta function are respectively replaced by the summation and the Kronecker's delta,
the quadratic nonlinear terms have the coefficient $2\pi$.

\subsection{Configuration of Numerical Simulations}
In our numerical simulations the wavenumber space $\bm{k} = (k_x,k_y)$
is discretized by the equally-distributed grids with the interval $\Delta k=1/42$ in both $k_x$ and $k_y$ directions
and is truncated along $|k_x|=k_{\rm max}$ and $|k_y|=k_{\rm max}$ with $k_{\rm max}=512\Delta k \approx 12$.
The convolutions in the nonlinear terms are obtained by the pseudospectral transform method.
In this method we first use inverse FFT of size $n_x=n_y=1024$ to transform $a_{\bm{k}}$ to
its inverse transform in the physical $\bm{x}$ space, perform there suitable multiplications,
and then use the FFT to obtain the convolution sums.
Although this pseudospectral transform method contains the alising error,
the region $|k_x|, |k_y| \leq 341\Delta k\approx 8$ in the $\bm{k}$ space is free from
this aliasing error due to the $3/2$-rule.
(For the $3/2$-rule, see for example \cite{Canuto}.)
We trace the temporal evolutions of $a_{\bm{k}}$ only for those $\bm{k}$'s which are within this alias-free region.

By reference to the Pierson--Moskowitz spectrum that is typical in the ocean waves,
we employ an isotropic spectrum as follows for the initial wave field:
\begin{subequations}
\begin{align}
& H_2=\sum_{\bm{k}}\omega_{\bm{k}}|a_{\bm{k}}|^2, \\
& |a_{\bm{k}}|^2=A  k^{-6.5} \exp(-1/k^4) D(k),
\label{eq:ainit}
 \\
& D(k)=
\begin{cases}
1, & (0<k<7), \\
\exp(-10(k-7)^2), & (7 \leq k \leq 8),
\end{cases}
\label{eqn:initial spectrum}
\end{align}
\label{eq:initsp}
\end{subequations}
where $H_2$ is the discrete counterpart of the lowest-order Hamiltonian $\mathcal{H}_2$ of (\ref{eqn:H2}).
The exponential function and the power-law function in Eq.~(\ref{eq:ainit}) respectively give
the increase in the small wavenumbers and the decrease in the large wavenumbers.
The function $D(k)$ gives the exponential tail near the end of the alias-free wavenumbers
so that the truncation in the $\bm{k}$ space does not affect the numerical results.
For the purpose of this study the choice of the spectrum is arbitrary,
and this spectrum (\ref{eq:initsp}) does not have any special significance
for the system at all.
Here, the coefficient $A$ is a parameter to control the value of $H_2$.
In this study,
we performed four series of simulations which have
$H_2 = 1.25 \times 10^{-6}$, $2.5 \times 10^{-6}$, $5 \times 10^{-6}$,
and $1 \times 10^{-5}$.
The initial phases of each component wave are given by uniform random numbers in the range $[0, 2\pi]$.
The ratio $|H_3/H_2|$ 
can be a measure of the degree of nonlinearity of the wave field as a whole,
where $H_3$ is the discrete counterpart of the interaction Hamiltonian $\mathcal{H}_3$ of (\ref{eqn:H3}).
$|H_3/H_2|$ is an increasing function of $H_2$, and takes values around
$5.0\times10^{-5}$ when $H_2=1.25\times10^{-6}$ and $3.5\times10^{-4}$ when $H_2=1\times10^{-5}$.
Our selection of the values of $H_2$ as above is made to keep $|H_3/H_2|$ and hence the nonlinearity of the wave field
 sufficiently small.
To make ensemble average,
$256$ independent simulations which have the different initial phases
are performed for each $H_2$.
The time integration is made until $t=100T_p$.
Here, $T_p=2\pi$ is the period given by the linear dispersion relation for $k=1$,
at which the one-dimensional energy spectrum defined below has its maximum.
The fourth-order Runge--Kutta method with the time interval $\Delta t = T_p/50$ is employed for the time integration.
The linear term is implicitly solved to improve the numerical stability.
Because the system and the initial spectrum are isotropic,
$a_{\bm{k}}$ for $k>8$ is set to $0$ at each time step.

The total Hamiltonian that is the sum of the linear part $H_2$ and the nonlinear part $H_3$ 
is numerically conserved within the relative error $2.6 \times 10^{-4}$
for $H_2 = 1 \times 10^{-5}$ where the nonlinearity is the largest
and the conservation is the worst.
In this case, the average of $H_3$ during $100T_p$ is $-3.5 \times 10^{-9}$.

All the numerical simulations are performed on FUJITSU FX1
in Information Technology Center, Nagoya University.
The CPU time for one realization takes 12 hours.

\section{Numerical Results}
\label{sec:numerical result}
\subsection{Time Evolution of Spectra}

\begin{figure*}
\includegraphics[scale=1]{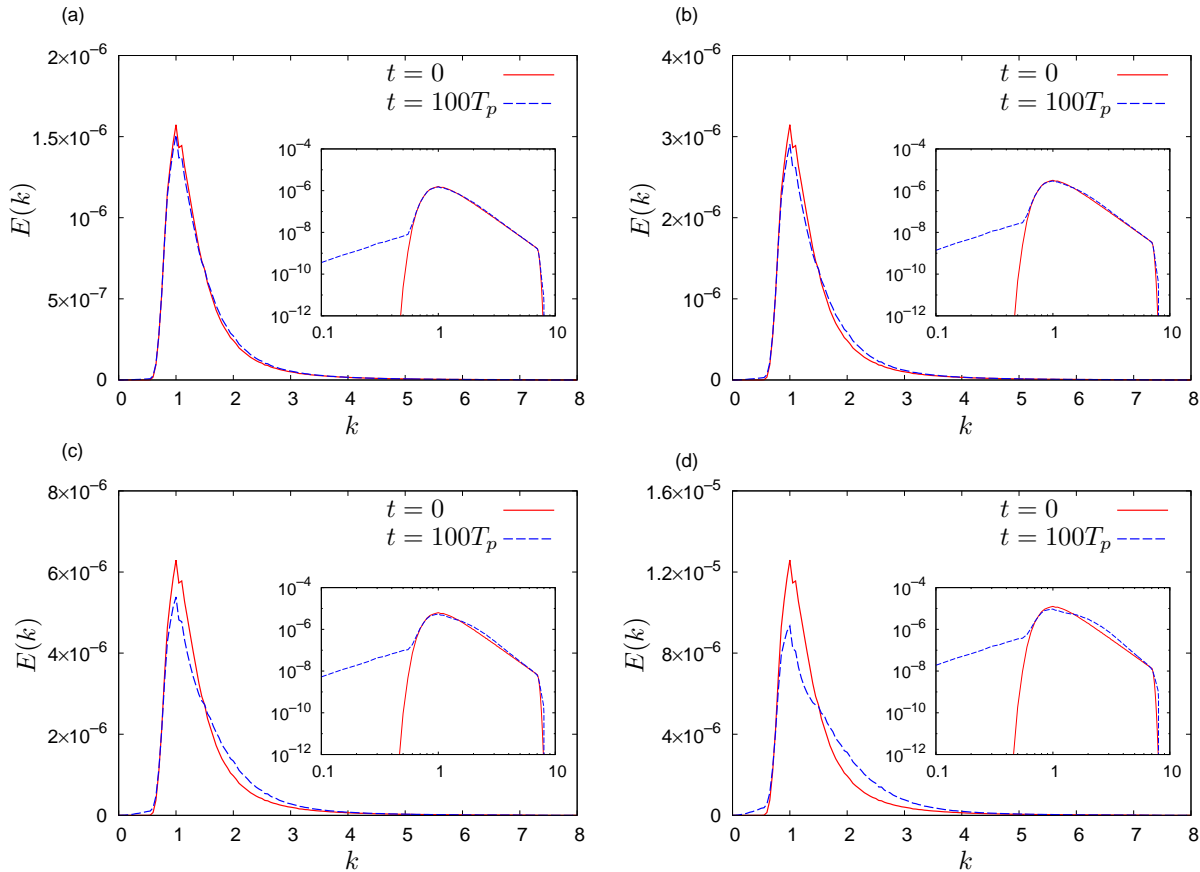}
\caption{(Color online)
Variations of one-dimensional energy spectra $E(k)$ from $t=0$ to $t=100T_p$.
The spectra are shown with the double logarithmic scales in the insets.
(a): $H_2=1.25 \times 10^{-6}$, (b): $H_2=2.5 \times 10^{-6}$,
(c): $H_2=5 \times 10^{-6}$, (d): $H_2=1 \times 10^{-5}$.
\label{fig:kspect1d}}
\end{figure*}

Figure~\ref{fig:kspect1d} shows the azimuthally-integrated one-dimensional energy spectra $E(k)$
\begin{equation}
E(k) = \frac{1}{\Delta_{\mathrm{bin}}} \sum_{k-\Delta_{\mathrm{bin}}/2 < |\bm{k}^{\prime}| < k+\Delta_{\mathrm{bin}}/2}
\langle \omega_{\bm{k}^{\prime}} |a_{\bm{k}^{\prime}}|^2 \rangle
,
\end{equation}
obtained from DNS at $t=100T_p$.
The initial spectra are also shown for reference.
Here, $\Delta_{\mathrm{bin}}$ denotes the width of the bins
which is used to evaluate $E(k)$
from the complex amplitudes defined on the discrete wavenumbers, and we set $\Delta_{\mathrm{bin}}=0.05$.
Figures~\ref{fig:kspect1d} (a), (b), (c) and (d)
show the energy spectra 
for $H_2 = 1.25 \times 10^{-6}$,
for $H_2 = 2.5 \times 10^{-6}$,
for $H_2 = 5 \times 10^{-6}$,
and for $H_2 = 1 \times 10^{-5}$, respectively.
Small irregularities around $k=1.1$ are
due to the numerical procedures to obtain $E(k)$
in the discrete $\bm{k}$ space.
While the variation of the spectrum during $100T_p$ for $H_2 = 1.25 \times 10^{-6}$ is quite small,
that for $H_2=1 \times 10^{-5}$ is large during the time.
Most of the weak turbulence theory, including RPA studied here, have
been developed for the {\em weak\/} turbulence,
where the linear time scale
determined by the linear frequency and the nonlinear time scale,
i.e., the time scale of spectral change are largely separated.
Therefore,
the weakly nonlinear assumption might be slightly violated for $H_2 = 1 \times 10^{-5}$,
although the assumption is evidently valid for $H_2 = 1.25 \times 10^{-6}$.

The wave action $n(\bm{k})$ of the three-wave system~(\ref{eq:dynamic}) is known to evolve according to the
following kinetic equation (See, for example, Refs.~\cite{zak_book,nazarenkobook}):
\begin{widetext}
\begin{subequations}
\begin{align}
\frac{d n(\bm{k})}{dt} &= -\gamma(\bm{k}) n(\bm{k})+\eta(\bm{k}), 
\label{eqn:kinetic equation}\\
\eta(\bm{k}) &= \pi \int
 \left(
\left|V(\bm{k}, \bm{k}_1, \bm{k}_2)\right|^2\delta(\bm{k} - \bm{k}_1 - \bm{k}_2)\delta(\omega(\bm{k}) - \omega(\bm{k}_1) - \omega(\bm{k}_2))
\right.
\nonumber\\
& \qquad\qquad\qquad
\left.
+2\left|V(\bm{k}_2, \bm{k}, \bm{k}_1)\right|^2
 \delta(\bm{k}_2 - \bm{k} - \bm{k}_1)
 \delta(\omega(\bm{k}_2) - \omega(\bm{k}) - \omega(\bm{k}_1))
\right)
n_{\bm{k}_1} n_{\bm{k}_2} 
d\bm{k}_{12}, 
\label{eqn:eta}\\
\gamma(\bm{k}) &= 2\pi \int \left(
\left|V(\bm{k}, \bm{k}_1, \bm{k}_2)\right|^2
 \delta(\bm{k} - \bm{k}_1 - \bm{k}_2)
 \delta(\omega(\bm{k}) - \omega(\bm{k}_1) - \omega(\bm{k}_2)) n(\bm{k}_2)
\right.
\nonumber\\
& \qquad\qquad\qquad
\left.
 +
 \left|V(\bm{k}_2,\bm{k},\bm{k}_1)\right|^2
 \delta(\bm{k}_2 - \bm{k} - \bm{k}_1)
 \delta(\omega(\bm{k}_2) - \omega(\bm{k}) - \omega(\bm{k}_1))
 (n(\bm{k}_2) - n(\bm{k}_1))\right) d\bm{k}_{12}.
\label{eqn:gamma}
\end{align}
\label{eq:kinetic}
\end{subequations}
\end{widetext}

\begin{figure}
\includegraphics[scale=1]{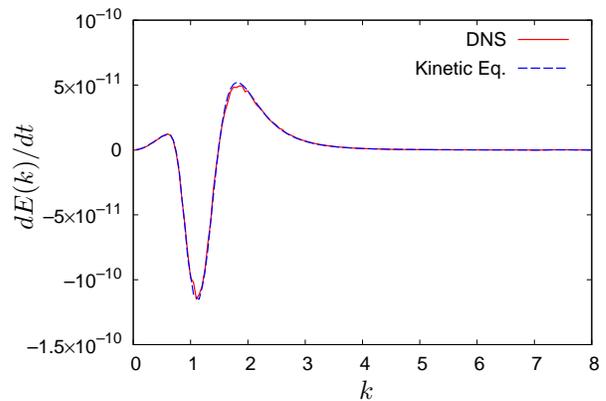}
\caption{(Color online)
Comparison of the time rates of change of $E(k)$ between DNS and the kinetic equation.
$H_2=1.25 \times 10^{-6}$.
\label{fig:dEdt_kinetic_DNS_Hs}}
\end{figure}

Since the energy spectrum for $H_2=1.25 \times 10^{-6}$ varies very little during the time $100T_p$ as shown in Fig.~\ref{fig:kspect1d}(a),
the right hand side of Eq.~(\ref{eqn:kinetic equation}) is almost constant in time
and the wave action is expected to be a linear function of time.
Figure~\ref{fig:dEdt_kinetic_DNS_Hs} shows
the time rates of change of $E(k)$ for $H_2=1.25 \times 10^{-6}$.
One is obtained for the initial spectrum~(\ref{eq:initsp})
according to Eq.~(\ref{eq:kinetic}).
The other is obtained from DNS
as the difference between the energy spectrum at $t=50T_p$ and that at $t=0$ divided by $50T_p$.
Both time rates of change agree quite well.

\begin{figure}
\includegraphics[scale=1]{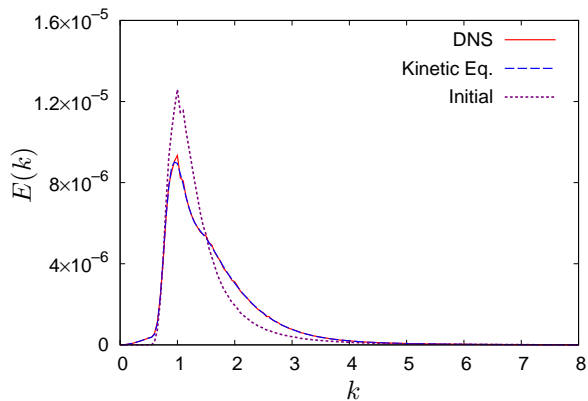}
\caption{(Color online)
Comparison of the variations of $E(k)$
from $t=0$ to $t=100T_p$
 between DNS and the kinetic equation.
 $H_2=1 \times 10^{-5}$.
\label{fig:dEdt_kinetic_DNS_Hl}}
\end{figure}

The same procedure of the comparison cannot be used for $H_2=1 \times 10^{-5}$,
since the variation of $E(k)$ is large.
Then,
the energy spectrum at $100T_p$ is obtained
by the numerical integration of Eq.~(\ref{eq:kinetic}) in time,
and it is compared with the energy spectrum at $t=100T_p$ obtained by DNS,
which is already shown in Fig.~\ref{fig:kspect1d}(d).
The comparison is shown in Fig.~\ref{fig:dEdt_kinetic_DNS_Hl}.
Also for $H_2=1 \times 10^{-5}$,
both spectra agree quite well.
It clearly shows
that the spectrum in DNS evolves in time
according to the prediction of the kinetic equation~(\ref{eq:kinetic}).

\subsection{Importance of Resonant Interaction for Evolution of Fluctuation}
In our DNS, the initial value of the amplitude of each wave mode $|a_{\bm{k}}(0)|$ is determined
by the initial spectrum,
hence no amplitude fluctuation exists at $t=0$.
The fluctuation grows as time elapses
through nonlinear interactions with other modes.
\begin{figure}
\includegraphics[scale=1]{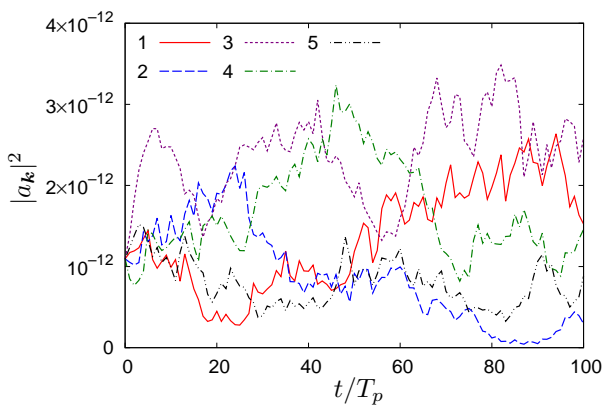}
\caption{(Color online)
Evolution of $|a_{\bm{k}}|^2$ in five independent realizations.
 $H_2 = 5 \times 10^{-6}$, $\bm{k}= (3,0)$.
\label{fig:noisiness}}
\end{figure}
Examples of the evolution of $|a_{\bm{k}}|^2$ are shown in Fig.~\ref{fig:noisiness}.
The five curves show the variations of $|a_{\bm{k}}|^2$ for $\bm{k}=(3,0)$
in five independent realizations for $H_2 = 5 \times 10^{-6}$.
Since the simulations are started without the amplitude fluctuations,
all $|a_{\bm{k}}|^2$ has the same value at $t=0$.
The fluctuations grow as time elapses,
and each $|a_{\bm{k}}|^2$ evolves differently.

\begin{figure*}
\includegraphics[scale=1]{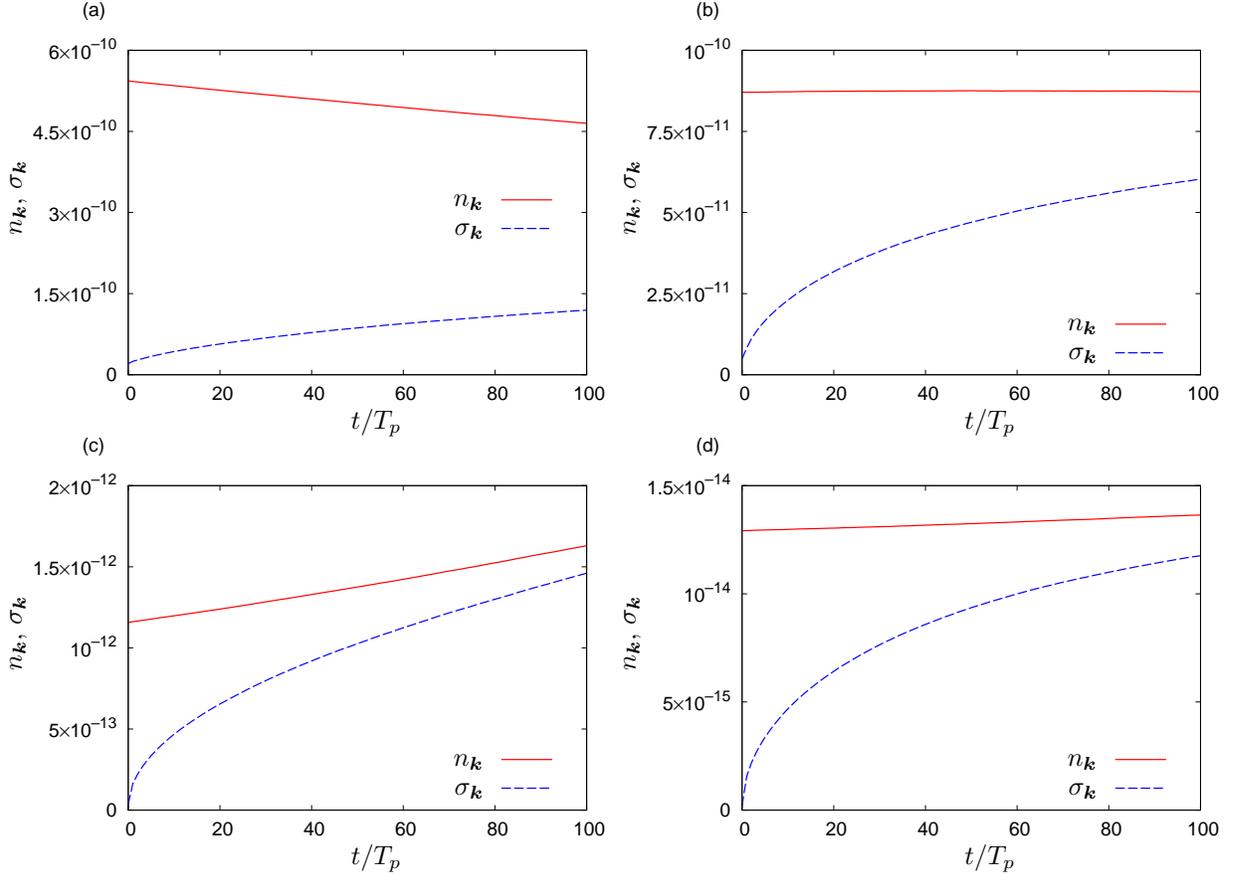}
\caption{(Color online)
Evolution of the means $n_{\bm{k}}$ and the standard deviations $\sigma_{\bm{k}}$
of $|a_{\bm{k}}|^2$.
$H_2=5 \times 10^{-6}$.
(a): $k=1$, (b): $k=1.5$, (c): $k=3$, (d): $k=6$.
\label{fig:s_and_sigma_capi}}
\end{figure*}

Figure~\ref{fig:s_and_sigma_capi} shows
the time evolutions of the mean $n_{\bm{k}}$ and the standard deviation $\sigma_{\bm{k}}$ of $|a_{\bm{k}}|^2$
obtained from DNS for $H_2 = 5 \times 10^{-6}$.
Figures~\ref{fig:s_and_sigma_capi} (a), (b), (c) and (d) respectively show the evolution of $n_{\bm{k}}$ and $\sigma_{\bm{k}}$
of $k=1$, $k=1.5$, $k=3$ and $k=6$.
Throughout this paper,
to evaluate the statistical quantities at $k$ such as $n_{\bm{k}}$,
the quantity is averaged over the modes of $\bm{k}^{\prime}$ in the annular domain $|k-k^{\prime}|<
\Delta_{\mathrm{bin}}/2\,(=0.025)$.
It can been seen that $\sigma_{\bm{k}}$, that is, the amplitude fluctuation grows in time at each $k$,
although the growth rates are wavenumber-dependent.
In particular, at the larger wavenumbers $k=3$ and $k=6$,
it is clearly observed that the fluctuations
are approaching the Gaussianity of $a_{\bm{k}}$, i.e.,  $n_{\bm{k}} = \sigma_{\bm{k}}$.

\begin{figure}
\includegraphics[scale=1]{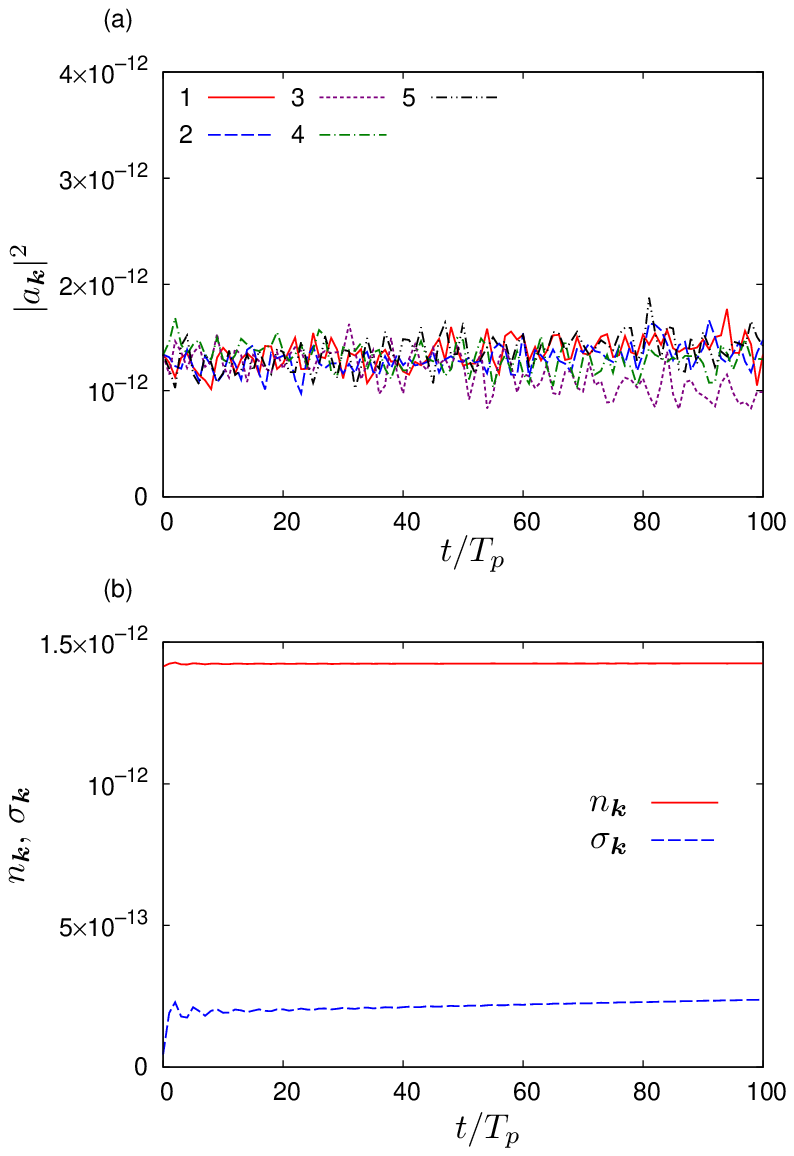}
\caption{(Color online)
Evolution of the fluctuations of $|a_{\bm{k}}|^2$.
$\omega=k^{1/2}$, $H_2=5 \times 10^{-6}$.
(a): $|a_{\bm{k}}|^2$ in five independent realizations.
 $\bm{k} = (3,0)$.
(b):
The mean $n_{\bm{k}}$ and the standard deviation $\sigma_{\bm{k}}$
of $|a_{\bm{k}}|^2$.
$k=3$.
\label{fig:s_and_sigma_gravi}}
\end{figure}

For comparison, we performed another series of DNS where the power-law exponent $\alpha$ of
the linear dispersion relation in Eq.~(\ref{eqn:model_2}) is changed to $\alpha = 1/2$ while
all the other aspects of the model remain intact.
In this case, the dispersion relation $\omega = k^{1/2}$ is of non-decay type,
which is similar to the surface gravity waves in water,
and the three-wave resonant interactions are prohibited.
The results are shown in Fig.~\ref{fig:s_and_sigma_gravi} where 
we draw the same quantities shown in Figs.~\ref{fig:noisiness} and \ref{fig:s_and_sigma_capi}(c).
By comparing Fig.~\ref{fig:s_and_sigma_gravi} with Figs.~\ref{fig:noisiness} and \ref{fig:s_and_sigma_capi}(c),
we observe that
the growth of the fluctuations is much slower
when the three-wave resonant interactions are prohibited by the dispersion relation.
Similarly,
the slower growths are observed for other wavenumbers.

The kinetic equation~(\ref{eq:kinetic}) derived by the weak turbulence theory
represents that
only the resonant interactions
play a role in the evolution of the wave action $n_{\bm{k}}$.
It can simply be understood
since the secular energy transfer persistent against the time average is necessary
for $n_{\bm{k}}$, which is the mean of $|a_{\bm{k}}|^2$, to evolve
and only the resonant interactions can provide it.
On the other hand,
the fluctuations of $|a_{\bm{k}}|^2$ around its mean
could grow
owing to the non-resonant nonlinear interactions as well as the resonant ones,
for the non-resonant interactions seem to work
as the stochastic driving forces in random walk processes. 
Against the intuition, however,
the comparison between Figs.~\ref{fig:noisiness} and \ref{fig:s_and_sigma_capi} and Fig.~\ref{fig:s_and_sigma_gravi} 
clearly shows that the resonant interactions are essential for the evolution of the fluctuations.
Even though the three-wave resonances are prohibited when $\alpha=1/2$ and our model
(\ref{eqn:continuous 3wave equation}) contains only quadratic nonlinear terms,
the fluctuations do grow as shown in Fig.~\ref{fig:s_and_sigma_gravi}(b) albeit very slowly.
This is because two non-resonant three-wave interactions can make a resonant four-wave interaction,
and hence the fluctuations grow due to this four-wave resonant interaction with a much slower time scale.

\subsection{Approach to Gaussianity}
The $p$th-order moment is defined as
\begin{equation}
M^{(p)}(\bm{k}) = \left\langle \left(\frac{|a(\bm{k})|^2}{\delta(\bm{0})}\right)^{p} \right\rangle
,
\end{equation}
where $\delta(\bm{0})$ is defined in the large-box limit
as
\begin{align}
\delta(\bm{0}) = \lim_{L_x, L_y \to \infty} \frac{L_x L_y}{(2\pi)^2}
= \lim_{\Delta k_x, \Delta k_y \to 0} \frac{1}{\Delta k_x \Delta k_y},
\end{align}
according to the correspondence (\ref{eqn:relation between delta}).
RPA predicts that $M^{(p)}(\bm{k})$ evolves according to the following equation~\cite{Choi2005361}:
\begin{equation}
\frac{d M^{(p)}(\bm{k})}{dt} =
 -p \gamma(\bm{k}) M^{(p)}(\bm{k}) + p^2 \eta(\bm{k}) M^{(p-1)}(\bm{k})
,
\label{eqn:equation for M}
\end{equation}
where $\eta(\bm{k})$ and $\gamma(\bm{k})$ are respectively given in Eq.~(\ref{eqn:eta}) and Eq.~(\ref{eqn:gamma}).
Equation~(\ref{eqn:equation for M}) for $p=1$
is identical to the kinetic equation~(\ref{eq:kinetic})
for the wave action $n(\bm{k})$.

When the real and imaginary parts of $a(\bm{k})$ are independent and obey the same Gaussian distribution,
\begin{equation}
M^{(p)}(\bm{k}) = p! \, n^p(\bm{k})
.
\end{equation}
Then,
\begin{equation}
F^{(p)}(\bm{k})=\frac{M^{(p)}(\bm{k}) - p! \, n^p(\bm{k})}{p! \, n^p(\bm{k})}
,
\end{equation}
is an index which expresses the deviation from the Gaussianity of $a(\bm{k})$.
Equation~(\ref{eqn:equation for M}) can be written for $F^{(p)}(\bm{k})$ as
\begin{equation}
\frac{d F^{(p)}(\bm{k})}{dt} = \frac{p \eta(\bm{k})}{n(\bm{k})}\left(F^{(p-1)}(\bm{k})-F^{(p)}(\bm{k})\right)
.
\label{eqn:equation for F}
\end{equation}
Because $F^{(1)}(\bm{k})=0$ by the definition that $M^{(1)}(\bm{k}) = n(\bm{k})$,
the equation for the evolution of $F^{(2)}(\bm{k})$ can be obtained as
\begin{equation}
\frac{d F^{(2)}(\bm{k})}{dt} = -\frac{2\eta(\bm{k})}{n(\bm{k})}F^{(2)}(\bm{k})
.
\label{eqn:equation for F2}
\end{equation}
When $|a(\bm{k})|$ is deterministically given and no fluctuation is allowed
like in the initial conditions of our DNS,
$F^{(2)}(\bm{k}) = -1/2$.
Equation~(\ref{eqn:equation for F}) represents that
the deviation from the Gaussianity of $a(\bm{k})$ for all $p$ decays after the elapse of sufficient time,
because $\eta(\bm{k})>0$.
The homogeneous term in Eq.~(\ref{eqn:equation for F})
indicates that
$F^{(p)}(\bm{k})$ for large $p$ decays fast,
and then the non-homogeneous term of the order $p-1$ can be dominant in the evolution of $F^{(p)}(\bm{k})$.
Therefore,
the evolution of $F^{(p)}(\bm{k})$ for all $p$ is determined by the slowest $F^{(2)}(\bm{k})$.

According to Eqs.~(\ref{eqn:equation for F}) and (\ref{eqn:equation for F2}),
the speed for $a(\bm{k})$ to approach the Gaussianity
depends on the value of $\eta(\bm{k})/n(\bm{k})$.
\begin{figure}
 \includegraphics[scale=1]{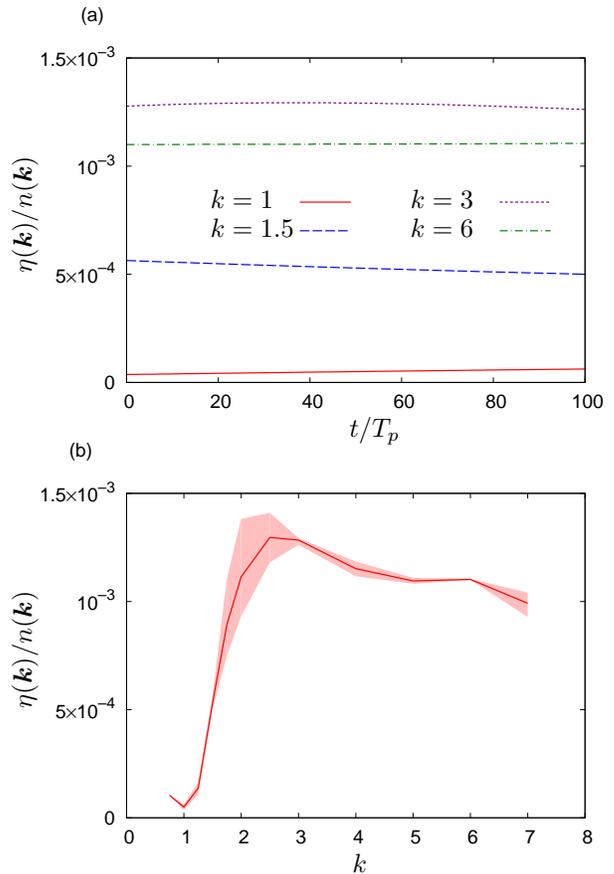}
\caption{(Color online)
Value of $\eta(\bm{k})/n(\bm{k})$ for each wavenumber.
 $H_2=5 \times 10^{-6}$.
 (a): Evolution,
 (b): time-average, maximum and minimum during $0 \leq t \leq 100T_p$.
 \label{fig:eta_over_n_H=5.00d-6}
 }
\end{figure}
Figure~\ref{fig:eta_over_n_H=5.00d-6}(a) shows
the evolution of $\eta(\bm{k})/n(\bm{k})$ for $k=1$, $1.5$, $3$ and $6$
for $H_2 = 5 \times 10^{-6}$,
and Fig.~\ref{fig:eta_over_n_H=5.00d-6}(b) shows the time-average, the maximum and the minimum of $\eta(\bm{k})/n(\bm{k})$ at each $k$
during $100T_p$.
These results are obtained by the numerical integration of the kinetic equation~(\ref{eq:kinetic}).
For the initial spectrum (\ref{eqn:initial spectrum}),
$\eta(\bm{k})/n(\bm{k})$ is large in the range $2 \lesssim k \lesssim 3$.
The wavenumber dependence of $\eta(\bm{k})/n(\bm{k})$
is consistent with the rapid growth of the fluctuation at $k=3$
compared with those at the other wavenumbers observed
in Fig.~\ref{fig:s_and_sigma_capi}.

As shown in Fig.~\ref{fig:kspect1d},
the spectral variation for $H_2 = 5 \times 10^{-6}$ during $100T_p$
is not small.
However,
Fig.~\ref{fig:eta_over_n_H=5.00d-6}(b) indicates that
the variation of the value of $\eta(\bm{k})/n(\bm{k})$ during $100T_p$
is almost constant in time except $1.5 < k < 3$,
because the differences between the maximums and the minimums are small.
Then,
Eq.~(\ref{eqn:equation for F2}) suggests that
$F^{(2)}(\bm{k})$ shows exponential decay in time.
\begin{figure}
 \includegraphics[scale=1]{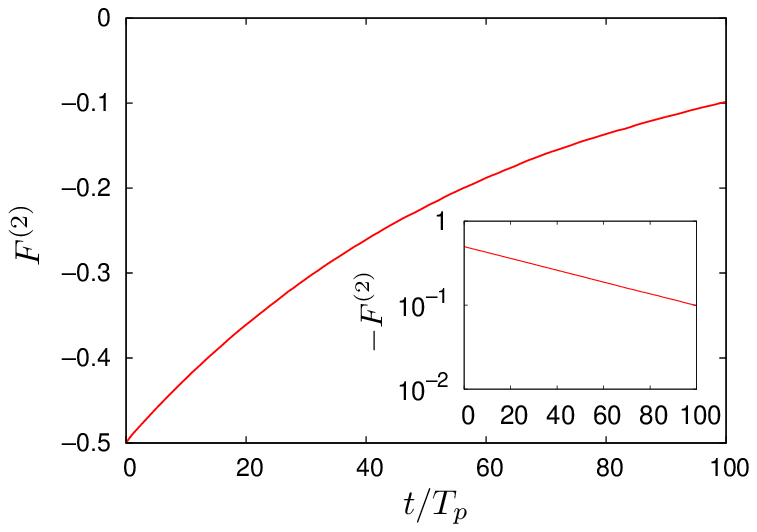}
\caption{(Color online)
Evolution of $F^{(2)}$ in DNS.
 $H_2 = 5 \times 10^{-6}$, $k=3$.
 The evolution is shown with the single logarithmic scale in the inset.
 \label{fig:F2_H=5.00d-6_k=3.00}
 }
\end{figure}
Figure~\ref{fig:F2_H=5.00d-6_k=3.00} shows the evolution of $F^{(2)}$ of $k=3$
for $H_2=5\times10^{-6}$ in DNS.
It is found that the absolute value of $F^{(2)}$ exponentially decays
as expected.
When $F^{(2)}$ is fitted by an exponential function $-1/2 \,\exp(-\lambda(\bm{k}) t)$,
the decay rate $\lambda(\bm{k})$ obtained by the method of least squares
is $2.58 \times 10^{-3}$.
On the other hand,
the decay rate predicted by RPA is $2\eta(\bm{k})/n(\bm{k})$.
The time-average of $2\eta(\bm{k})/n(\bm{k})$ during $100T_p$
is $2.57 \times 10^{-3}$ for $k=3$.
Thus the quantitative agreement between RPA and DNS is quite good.

\begin{figure}
 \includegraphics{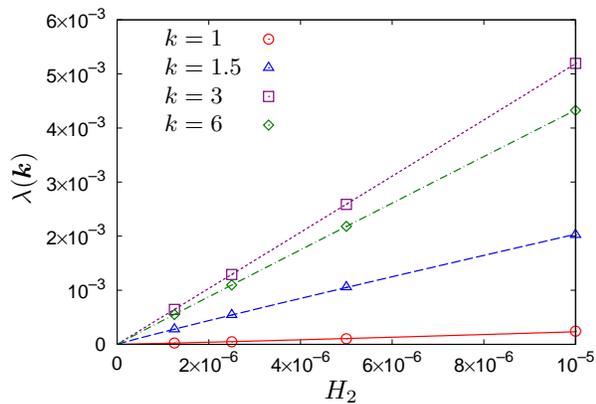}
 \caption{(Color online)
Dependence of the decay rate $\lambda(\bm{k})$ of $F^{(2)}(\bm{k})$
 in DNS
 on $H_2$.
 \label{fig:alpha_of_F2_vs_H2}
}
\end{figure}
Figure~\ref{fig:alpha_of_F2_vs_H2} shows
the decay rate $\lambda(\bm{k})$ for $k=1$, $1.5$, $3$ and $6$
obtained by the least-square fit of the exponential function to $F^{(2)}(\bm{k})$ in DNS
as a function of $H_2$.
It is observed that
$\lambda(\bm{k})$ for each wavenumber is proportional to $H_2$.
As shown by Eqs.~(\ref{eq:kinetic}) and (\ref{eqn:equation for F2}),
RPA predicts that
the time scales of the growth of the fluctuations as well as those of the evolution of the spectra
are inversely proportional to the value of the Hamiltonian.
It is consistent to the results shown in Fig.~\ref{fig:alpha_of_F2_vs_H2}.

\subsection{Moments for $p>2$}
The solution of Eq.~(\ref{eqn:equation for F})
is given as
\begin{align}
& F^{(p)}(\bm{k},t) = \sum_{j=2}^p C_j^{(p)} e^{-j\theta(\bm{k})},
 \nonumber\\
& \theta(\bm{k}) = \int_0^t \frac{\eta(\bm{k}, t^{\prime})}{n(\bm{k}, t^{\prime})}\, dt^{\prime}
.
\label{eqn:solution of Fp}
\end{align}
The coefficient $C_j^{(p)}$ is determined by the recurrence formula
\begin{align}
C_j^{(p)}=
\left(
\begin{array}{c}
p \\
j
\end{array}
\right)
C_j^{(j)} \quad (j=2,\ldots,p-1),
\nonumber\\
C_2^{(2)}=F^{(2)}(\bm{k},0),
\quad
C_p^{(p)}=F^{(p)}(\bm{k},0)-\sum_{j=2}^{p-1} C_j^{(p)}
,
\label{eqn:coefficient for Fp}
\end{align}
where
$\left(
\begin{array}{c}
p  \\
j
\end{array}
\right)
$
is the binomial coefficient~\footnote{
The explicit expression of the solution of Eq.~(\ref{eqn:equation for F})
in Ref.~\cite{lvov2004nsl} is incorrect.
}.

\begin{figure}
 \includegraphics[scale=1]{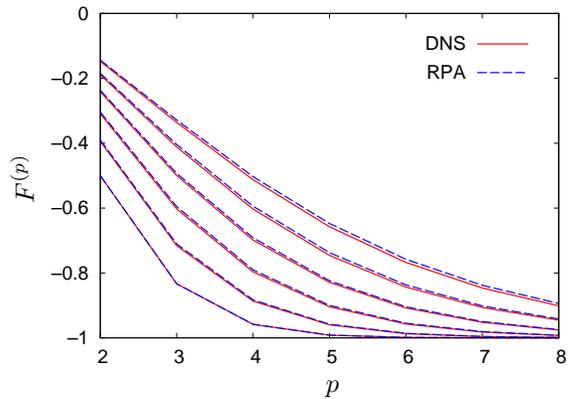}
\caption{(Color online)
Comparison of $F^{(p)}$ between DNS and RPA for each $p$.
 $t=0, 20T_p, \cdots, 100T_p$ from bottom to top.
 $H_2 = 5 \times 10^{-6}$, $k=7$.
\label{fig:Fp(t)_vs_p_RPA_DNS}
}
\end{figure}

\begin{figure}
\includegraphics[scale=1]{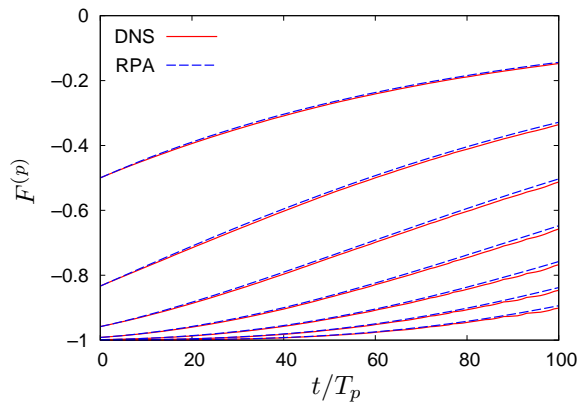}
\caption{(Color online)
Evolution of $F^{(p)}$ in DNS and RPA.
 $p=2, 3, \cdots, 8$ from top to bottom.
 $H_2 = 5 \times 10^{-6}$, $k=7$.
\label{fig:F5_F10_RPA_DNS}
}
\end{figure}

The deviations from the Gaussianity $F^{(p)}$ at $k=7$
at every $20T_p$ from $t=0$ to $t=100T_p$
for $H_2 = 5 \times 10^{-6}$
obtained by Eq.~(\ref{eqn:solution of Fp})
and DNS
are shown in Fig.~\ref{fig:Fp(t)_vs_p_RPA_DNS}
as functions of the order $p$ of the moments.
A large number of $a_{\bm{k}}$
is required to obtain reliable high-order moments.
As explained before, to evaluate statistical quantities such as the moments of $a_{\bm{k}}$ at $k$,
we use the values of $a_{\bm{k'}}$ in the annular domain $|k'-k|<0.025$,
hence the number of modes used in the statistical evaluation increases in proportion to $k$.
This gives the reason to choose the large wavenumber $k=7$ to evaluate the high-order moments.
We confirmed that
the tail of the probability density function of $|a_{\bm{k}}|^2$ defined below
decays faster than the negative $9$th power of $|a_{\bm{k}}|^2$.
Therefore, the moments up to the $8$th order are reliable.
To obtain the deviation $F^{(p)}$ by Eq.~(\ref{eqn:solution of Fp}) in RPA,
$\theta(\bm{k})$ is approximated by $\theta(\bm{k}) \approx \overline{(\eta/n)} t$,
where $\overline{(\eta/n)}$ is the time-averaged value of $\eta(\bm{k})/n(\bm{k})$ during $100T_p$.
Similarly,
the evolution of the deviations $F^{(p)}$ at $k=7$
obtained by DNS and RPA
are compared in Fig.~\ref{fig:F5_F10_RPA_DNS}.
Even though the value of $\theta(\bm{k})$ is approximated by $\overline{(\eta/n)} t$ 
and the high-order moments such as of $8$th-order
are treated,
RPA and DNS show another good agreement.

\subsection{Evolution of Distribution of Amplitude Fluctuations}
\begin{figure}
 \includegraphics[scale=1]{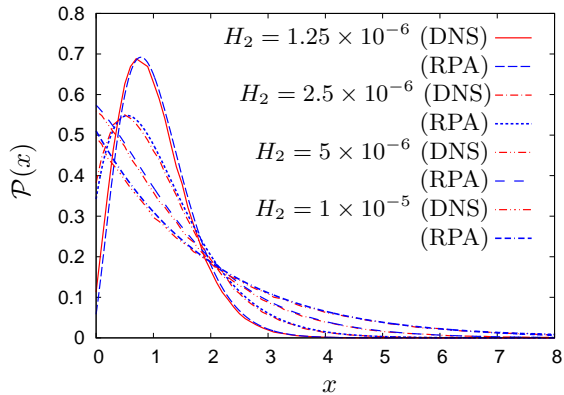}
 \caption{(Color online)
Comparison of $\mathcal{P}(x)$ in DNS and RPA. $k=3$, $t=100T_p$.
 \label{fig:P(s)_RPA_DNS}
}
\end{figure}

RPA also gives
the evolution of the probability density function (PDF) $\mathcal{P}(s(\bm{k}))$ of $s(\bm{k})$,
which is a stochastic variable defined by $s(\bm{k}) = |a(\bm{k})|^2 / \delta(\bm{0})$,
as follows~\cite{Choi2005361}:
\begin{align}
&
\frac{\partial \mathcal{P}(s(\bm{k}))}{\partial t}
 = \frac{\partial \mathcal{F}(\bm{k})}{\partial s(\bm{k})},
\nonumber\\
&
\mathcal{F}(\bm{k}) = s(\bm{k})
\left( \gamma(\bm{k}) \mathcal{P}(s(\bm{k}))
 + \eta(\bm{k}) \frac{\partial \mathcal{P}(s(\bm{k}))}{\partial s(\bm{k})} \right),
\label{eq:Ps}
\end{align}
where 
$\eta(\bm{k})$ and $\gamma(\bm{k})$
are the coefficients (\ref{eqn:eta}) and (\ref{eqn:gamma})
in the kinetic equation~(\ref{eq:kinetic}).
The PDF $\mathcal{P}(s)$ at $k=3$ for four values of $H_2$ is shown in Fig.~\ref{fig:P(s)_RPA_DNS}.
The PDF is obtained by the numerical integration of Eq.~(\ref{eq:Ps}) until $100T_p$.
Figure~\ref{fig:P(s)_RPA_DNS} has $s$ normalized by its initial value $n(0)$,
i.e., $x=s/n(0)$ on the abscissa.
The variation of $\mathcal{P}(x)$ is faster for larger $H_2$.
For $H_2 = 1.25 \times 10^{-6}$,
$\mathcal{P}(x)$ has a narrow distribution with a remnant of the initial distribution $\delta(x-1)$ even at $=100T_p$,
while for $H_2 = 1 \times 10^{-5}$, it has almost reached the $\chi^2$-distribution with $2$ degrees of freedom
corresponding to the Gaussianity of $a(\bm{k})$ by the same time.
The PDF obtained in DNS at $100T_p$
is also shown in Fig.~\ref{fig:P(s)_RPA_DNS}.
For all $H_2$,
the theoretical prediction of RPA and the result of DNS agree quite well.
RPA does not reply on any proximity to the Gaussianity of $a_{\bm{k}}$~\cite{lvov2004nsl}.
Figures~\ref{fig:Fp(t)_vs_p_RPA_DNS}, \ref{fig:F5_F10_RPA_DNS} and \ref{fig:P(s)_RPA_DNS}
confirm the validity of RPA
even when the Gaussianity of $a_{\bm{k}}$ has not been established at all.

The coefficients $\eta(\bm{k})$ and $\gamma(\bm{k})$ in Eq.~(\ref{eq:Ps})
vary as the wave action $n(\bm{k})$ evolves.
Then, in the calculation of Eq.~(\ref{eq:Ps}),
the calculation of the kinetic equation~(\ref{eqn:kinetic equation}) is first performed
to obtain $\gamma(\bm{k},t)$ and $\eta(\bm{k},t)$ as functions of time.
Moreover,
since $s(\bm{k})$ does not have fluctuations initially in this study,
$\mathcal{P}(x,0)=\delta(x-1)$ should be employed for the initial condition
to numerically obtain $\mathcal{P}(s)$ at later time according to Eq.~(\ref{eq:Ps}).
However, it causes numerical difficulty.
If $\mathcal{P}(x)$ is very narrow,
the diffusion term is dominant in the right-hand side of Eq.~(\ref{eq:Ps}).
Then,
in the very early stage of the evolution,
Eq.~(\ref{eq:Ps}) can be approximated by the diffusion equation
\begin{equation}
\frac{\partial \mathcal{P}(x)}{\partial t} = 
 \mu \frac{\partial^2 \mathcal{P}(x)}{\partial x^2},
\quad
\mu = \frac{x_0\eta_0}{n_0},
\quad
x_0 = 1,
\label{eqn:diffusion equation}
\end{equation}
where $\eta_0$ is $\eta$ at $t=0$.
This makes it possible that 
the original initial condition
$\mathcal{P}(x,0)=\delta(x-1)$
is replaced by the state after the PDF evolves according to Eq.~(\ref{eqn:diffusion equation})
for small time $t_0$,
i.e., the normal distribution with small standard deviation $\sigma_0$
\begin{equation}
\mathcal{P}(x,t_0) = 
 \frac{1}{\sqrt{2\pi}\sigma_0} \exp\left(-\frac{(x-1)^2}{2{\sigma_0}^2}\right),
\quad
(t_0 = \sigma_0^2/2\mu)
\end{equation}
to numerically solve the initial-value problem of Eq.~(\ref{eq:Ps}).
For the result in Fig.~\ref{fig:P(s)_RPA_DNS},
$\sigma_0=0.1$ is used.
Even if $\sigma_0=0.05$ is used,
perceptible change in Fig.~\ref{fig:P(s)_RPA_DNS} is not produced.
To obtain $\mathcal{P}(x)$ at $k=3$ in DNS,
$s_{\bm{k}}$ of $1,680$ modes in the annular domain $|k-3|<0.025$
are used.
Since $256$ realizations in DNS are independently performed,
$430,080$($=1,680 \times 256$) data is used to draw $\mathcal{P}(x)$ in Fig.~\ref{fig:P(s)_RPA_DNS}.

\section{Concluding Remark}
\label{sec:concludingremark}
In this work,
direct numerical simulations (DNS) for a three-wave resonant Hamiltonian system are performed,
and the validity of the Random Phase and Amplitude Formalism (RPA)
recently proposed in the weak turbulence
is evaluated by quantitative comparison with DNS.
It is confirmed that
the theoretical prediction of RPA and the result of DNS agree quite well
in all the statistical aspects of the amplitude fluctuations
such as the high-order moments and
the evolution of the probability density function (PDF) of $s(\bm{k}) = |a(\bm{k})|^2/\delta(\bm{0})$ of each mode,
the asymptotic approach of $a(\bm{k})$ to the Gaussianity and its time scales, and so on.
We have performed the comparison between RPA and DNS for the same three-wave model but
in the case of an anisotropic initial spectrum,
and obtained the same level of agreement between RPA and DNS. (Not shown here)

The comparison is restricted to the three-wave resonant system in this work.
In RPA for four-wave systems,
there exist some characteristics which the three-wave systems do not have
such as the renormalization of the nonlinear frequency.
Hence, the comparison for four-wave systems are still necessary.
The authors numerically investigated the growth of the amplitude fluctuations
in a four-wave system~\cite{MT_NY_physd}.
It was found that the time scales of the growth of the fluctuations
is much shorter than those of the spectral variations
in the large wavenumbers away from the spectral peak.
Although we did not attempt quantitative comparison with RPA at that time,
we felt suspicion about the validity of RPA because RPA predicts that
both time scales should be of the same order in terms of the amplitude expansion
or in terms of the Hamiltonian $H$.
This suspicion was actually one of the motives of the present work.
Judging from the complete agreement between RPA and DNS regarding to a
three-wave system reported here, we expect that this suspicion which we previously felt
about the validity of RPA would be cleared up when we will have finished the same kind
of comparison between RPA and DNS for some four-wave systems as well.
In Ref.~\cite{MT_NY_physd},
we also investigated statistical nature of $d s(\bm{k})/dt$ by DNS.
We numerically showed that
the fluctuation of $d s(\bm{k})/dt$ approaches a quasi-steady state
faster than $s(\bm{k})$
and analytically showed that
PDF of $d s(\bm{k})/dt$ has the Laplace distribution.
No work for $d s(\bm{k})/dt$ were found in the framework of RPA.
This is also one of our future work.

RPA gives theoretical prediction about the multi-mode statistics
such as the joint probability density $P^{(N)}$ of $|a_{\bm{k}_1}|^2, \ldots, |a_{\bm{k}_N}|^2$
as well as the single-mode statistics ~\cite{Choi2005121},
but we have restricted our attention in the present study only to the single-mode statistics
and have not treated those multi-mode statistics at all.
With regard to the multi-mode statistics, it is pointed out that the original
results by RPA are not correct~\cite{Eyink20121487}.
Then, numerical investigation of the multi-mode statistics remains to be done.

Our present numerical results also prompt a new question of the time scales
which is required by the weak turbulence theory.
\citeauthor{janssen_freak}~\cite{janssen_freak} derived
a kinetic equation in a different form
from that by \citeauthor{hasselmann}~\cite{hasselmann} and by \citeauthor{zak_book}~\cite{zak_book}.
He pointed out that
the conventional kinetic equation is recovered
by replacing the ``resonance function''
\begin{equation}
R_i(\Delta\omega,t) = \frac{\sin(\Delta\omega t)}{\Delta\omega}
,
\end{equation}
by the delta function which is the asymptotic of $R_i$
in the limit $t \to \infty$.
He claimed that
non-resonant interactions as well as resonant interactions
contribute the spectral evolution
until the time when the replacement is validated.
In RPA investigated here,
the important equations such as for the moments Eq.~(\ref{eqn:equation for M})
and for the PDF Eq.~(\ref{eq:Ps})
are derived.
It is based on the premise that
the linear time scale $\tau_l = O(2\pi/\omega)$
and the nonlinear time scale, i.e., the time scale of the spectral evolution $\tau_n = O(1/(\epsilon^2 \omega))$
are well separated,
and the intermediate time scale
\begin{equation}
\tau_l \ll \tau_i \ll \tau_n
,
\end{equation}
should exist.
Based on their derivation,
the equations in RPA are supposed to describe the time rate of change of the moments and the PDF
in the time scale of $\tau_i$.

\begin{figure}
 \includegraphics{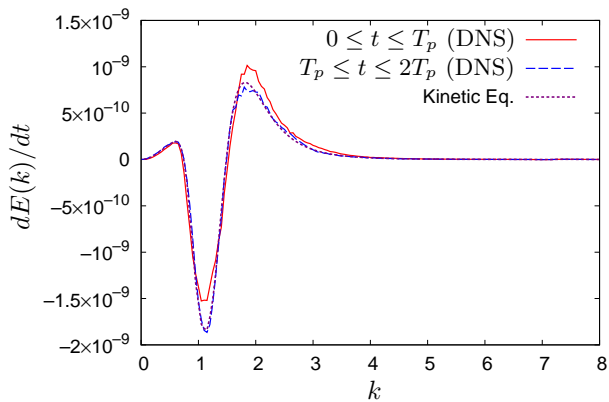}
\caption{(Color online)
Comparison of $dE(k)/dt$ obtained from the spectral variation during one period in DNS
and from kinetic equation.
\label{fig:dEdt_DNS_kineitc_short}
}
\end{figure}
In Fig.~\ref{fig:dEdt_DNS_kineitc_short},
the time rates of spectral changes $dE(k)/dt$ during $T_p$ from $t=0$ to $t=T_p$ and from $t=T_p$ to $t=2T_p$
for $H_2 = 5 \times 10^{-6}$
are compared with that obtained by the kinetic equation~(\ref{eq:kinetic})
for the initial spectrum.
Although the time rate of the spectral change for the first $T_p$
is slightly different from the prediction of the kinetic equation,
the rate during the period $T_p$ from $t=T_p$ to $t=2T_p$
almost perfectly coincides with the prediction of the kinetic equation.
The time rates of change of the higher-order moments during $T_p$
also show similar agreements between DNS and RPA.
These agreements between DNS and RPA in the short-time evolution
appears to contradict the procedure of the derivation of the statistical equations in RPA.
This short-time agreement has already been pointed out previously~\cite{doi:10.1175/JPO3029.1}.
This might affect the basis of the weak turbulence theory including RPA,
and must be investigated further.

\begin{acknowledgments}
The authors are grateful to Dr.\ Yuri V.\ Lvov and Dr.\ Yeontaek Choi for valuable discussions.
The numerical simulations are performed at Information Technology Center, Nagoya University.
\end{acknowledgments}


\begin{thebibliography}{23}%
\makeatletter
\providecommand \@ifxundefined [1]{%
 \@ifx{#1\undefined}
}%
\providecommand \@ifnum [1]{%
 \ifnum #1\expandafter \@firstoftwo
 \else \expandafter \@secondoftwo
 \fi
}%
\providecommand \@ifx [1]{%
 \ifx #1\expandafter \@firstoftwo
 \else \expandafter \@secondoftwo
 \fi
}%
\providecommand \natexlab [1]{#1}%
\providecommand \enquote  [1]{``#1''}%
\providecommand \bibnamefont  [1]{#1}%
\providecommand \bibfnamefont [1]{#1}%
\providecommand \citenamefont [1]{#1}%
\providecommand \href@noop [0]{\@secondoftwo}%
\providecommand \href [0]{\begingroup \@sanitize@url \@href}%
\providecommand \@href[1]{\@@startlink{#1}\@@href}%
\providecommand \@@href[1]{\endgroup#1\@@endlink}%
\providecommand \@sanitize@url [0]{\catcode `\\12\catcode `\$12\catcode
  `\&12\catcode `\#12\catcode `\^12\catcode `\_12\catcode `\%12\relax}%
\providecommand \@@startlink[1]{}%
\providecommand \@@endlink[0]{}%
\providecommand \url  [0]{\begingroup\@sanitize@url \@url }%
\providecommand \@url [1]{\endgroup\@href {#1}{\urlprefix }}%
\providecommand \urlprefix  [0]{URL }%
\providecommand \Eprint [0]{\href }%
\providecommand \doibase [0]{http://dx.doi.org/}%
\providecommand \selectlanguage [0]{\@gobble}%
\providecommand \bibinfo  [0]{\@secondoftwo}%
\providecommand \bibfield  [0]{\@secondoftwo}%
\providecommand \translation [1]{[#1]}%
\providecommand \BibitemOpen [0]{}%
\providecommand \bibitemStop [0]{}%
\providecommand \bibitemNoStop [0]{.\EOS\space}%
\providecommand \EOS [0]{\spacefactor3000\relax}%
\providecommand \BibitemShut  [1]{\csname bibitem#1\endcsname}%
\let\auto@bib@innerbib\@empty
\bibitem [{\citenamefont {Hasselmann}(1962)}]{hasselmann}%
  \BibitemOpen
  \bibfield  {author} {\bibinfo {author} {\bibfnamefont {K.}~\bibnamefont
  {Hasselmann}},\ }\href@noop {} {\bibfield  {journal} {\bibinfo  {journal} {J.
  Fluid Mech.}\ }\textbf {\bibinfo {volume} {12}},\ \bibinfo {pages} {481}
  (\bibinfo {year} {1962})}\BibitemShut {NoStop}%
\bibitem [{\citenamefont {Zakharov}\ and\ \citenamefont
  {Filonenko}(1967)}]{zak_filo1967}%
  \BibitemOpen
  \bibfield  {author} {\bibinfo {author} {\bibfnamefont {V.~E.}\ \bibnamefont
  {Zakharov}}\ and\ \bibinfo {author} {\bibfnamefont {N.~N.}\ \bibnamefont
  {Filonenko}},\ }\href@noop {} {\bibfield  {journal} {\bibinfo  {journal}
  {Sov. Phys. Dokl.}\ }\textbf {\bibinfo {volume} {10}},\ \bibinfo {pages}
  {881} (\bibinfo {year} {1967})}\BibitemShut {NoStop}%
\bibitem [{\citenamefont {Lvov}\ \emph {et~al.}(2003)\citenamefont {Lvov},
  \citenamefont {Nazarenko},\ and\ \citenamefont {West}}]{Lvov2003333}%
  \BibitemOpen
  \bibfield  {author} {\bibinfo {author} {\bibfnamefont {Y.}~\bibnamefont
  {Lvov}}, \bibinfo {author} {\bibfnamefont {S.}~\bibnamefont {Nazarenko}}, \
  and\ \bibinfo {author} {\bibfnamefont {R.}~\bibnamefont {West}},\ }\href@noop
  {} {\bibfield  {journal} {\bibinfo  {journal} {Physica D}\ }\textbf {\bibinfo
  {volume} {184}},\ \bibinfo {pages} {333} (\bibinfo {year}
  {2003})}\BibitemShut {NoStop}%
\bibitem [{\citenamefont {Connaughton}\ \emph {et~al.}(2003)\citenamefont
  {Connaughton}, \citenamefont {Nazarenko},\ and\ \citenamefont
  {Newell}}]{Connaughton200386}%
  \BibitemOpen
  \bibfield  {author} {\bibinfo {author} {\bibfnamefont {C.}~\bibnamefont
  {Connaughton}}, \bibinfo {author} {\bibfnamefont {S.}~\bibnamefont
  {Nazarenko}}, \ and\ \bibinfo {author} {\bibfnamefont {A.~C.}\ \bibnamefont
  {Newell}},\ }\href@noop {} {\bibfield  {journal} {\bibinfo  {journal}
  {Physica D}\ }\textbf {\bibinfo {volume} {184}},\ \bibinfo {pages} {86}
  (\bibinfo {year} {2003})}\BibitemShut {NoStop}%
\bibitem [{\citenamefont {Kartashova}(2010)}]{kartashova2010nonlinear}%
  \BibitemOpen
  \bibfield  {author} {\bibinfo {author} {\bibfnamefont {E.}~\bibnamefont
  {Kartashova}},\ }\href@noop {} {\emph {\bibinfo {title} {Nonlinear resonance
  analysis: theory, computation, applications}}}\ (\bibinfo  {publisher}
  {Cambridge Univ. Pr.},\ \bibinfo {year} {2010})\BibitemShut {NoStop}%
\bibitem [{\citenamefont {Newell}\ \emph {et~al.}(2001)\citenamefont {Newell},
  \citenamefont {Nazarenko},\ and\ \citenamefont {Biven}}]{newell01}%
  \BibitemOpen
  \bibfield  {author} {\bibinfo {author} {\bibfnamefont {A.~C.}\ \bibnamefont
  {Newell}}, \bibinfo {author} {\bibfnamefont {S.}~\bibnamefont {Nazarenko}}, \
  and\ \bibinfo {author} {\bibfnamefont {L.}~\bibnamefont {Biven}},\
  }\href@noop {} {\bibfield  {journal} {\bibinfo  {journal} {Physica D}\
  }\textbf {\bibinfo {volume} {152--153}},\ \bibinfo {pages} {520} (\bibinfo
  {year} {2001})}\BibitemShut {NoStop}%
\bibitem [{\citenamefont {Dyachenko}\ \emph {et~al.}(1992)\citenamefont
  {Dyachenko}, \citenamefont {Newell}, \citenamefont {Pushkarev},\ and\
  \citenamefont {Zakharov}}]{nls_intermittent_cycle}%
  \BibitemOpen
  \bibfield  {author} {\bibinfo {author} {\bibfnamefont {S.}~\bibnamefont
  {Dyachenko}}, \bibinfo {author} {\bibfnamefont {A.~C.}\ \bibnamefont
  {Newell}}, \bibinfo {author} {\bibfnamefont {A.}~\bibnamefont {Pushkarev}}, \
  and\ \bibinfo {author} {\bibfnamefont {V.~E.}\ \bibnamefont {Zakharov}},\
  }\href@noop {} {\bibfield  {journal} {\bibinfo  {journal} {Physica D}\
  }\textbf {\bibinfo {volume} {57}},\ \bibinfo {pages} {96} (\bibinfo {year}
  {1992})}\BibitemShut {NoStop}%
\bibitem [{\citenamefont {Newell}\ and\ \citenamefont
  {Rumpf}(2011)}]{ISI:000287046400003}%
  \BibitemOpen
  \bibfield  {author} {\bibinfo {author} {\bibfnamefont {A.~C.}\ \bibnamefont
  {Newell}}\ and\ \bibinfo {author} {\bibfnamefont {B.}~\bibnamefont {Rumpf}},\
  }\href@noop {} {\bibfield  {journal} {\bibinfo  {journal} {Annu. Rev. Fluid
  Mech.}\ }\textbf {\bibinfo {volume} {43}},\ \bibinfo {pages} {59} (\bibinfo
  {year} {2011})}\BibitemShut {NoStop}%
\bibitem [{\citenamefont {Pushkarev}(1999)}]{pushkarev_1999}%
  \BibitemOpen
  \bibfield  {author} {\bibinfo {author} {\bibfnamefont {A.~N.}\ \bibnamefont
  {Pushkarev}},\ }\href@noop {} {\bibfield  {journal} {\bibinfo  {journal}
  {Eur. J. Mech. B/Fluids}\ }\textbf {\bibinfo {volume} {18}},\ \bibinfo
  {pages} {345} (\bibinfo {year} {1999})}\BibitemShut {NoStop}%
\bibitem [{\citenamefont {D{\"{u}}ring}\ \emph {et~al.}(2006)\citenamefont
  {D{\"{u}}ring}, \citenamefont {Josserand},\ and\ \citenamefont
  {Rica}}]{during2006weak}%
  \BibitemOpen
  \bibfield  {author} {\bibinfo {author} {\bibfnamefont {G.}~\bibnamefont
  {D{\"{u}}ring}}, \bibinfo {author} {\bibfnamefont {C.}~\bibnamefont
  {Josserand}}, \ and\ \bibinfo {author} {\bibfnamefont {S.}~\bibnamefont
  {Rica}},\ }\href@noop {} {\bibfield  {journal} {\bibinfo  {journal} {Phys.
  Rev. Lett.}\ }\textbf {\bibinfo {volume} {97}},\ \bibinfo {pages} {025503}
  (\bibinfo {year} {2006})}\BibitemShut {NoStop}%
\bibitem [{\citenamefont {Zakharov}\ \emph {et~al.}(1992)\citenamefont
  {Zakharov}, \citenamefont {L'vov},\ and\ \citenamefont
  {Falkovich}}]{zak_book}%
  \BibitemOpen
  \bibfield  {author} {\bibinfo {author} {\bibfnamefont {V.~E.}\ \bibnamefont
  {Zakharov}}, \bibinfo {author} {\bibfnamefont {V.~S.}\ \bibnamefont {L'vov}},
  \ and\ \bibinfo {author} {\bibfnamefont {G.}~\bibnamefont {Falkovich}},\
  }\href@noop {} {\emph {\bibinfo {title} {{K}olmogorov Spectra of Turbulence
  I}}}\ (\bibinfo  {publisher} {Springer-Verlag},\ \bibinfo {address}
  {Berlin},\ \bibinfo {year} {1992})\BibitemShut {NoStop}%
\bibitem [{\citenamefont {Lvov}\ and\ \citenamefont
  {Nazarenko}(2004)}]{lvov2004nsl}%
  \BibitemOpen
  \bibfield  {author} {\bibinfo {author} {\bibfnamefont {Y.~V.}\ \bibnamefont
  {Lvov}}\ and\ \bibinfo {author} {\bibfnamefont {S.}~\bibnamefont
  {Nazarenko}},\ }\href@noop {} {\bibfield  {journal} {\bibinfo  {journal}
  {Phys. Rev. E}\ }\textbf {\bibinfo {volume} {69}},\ \bibinfo {pages} {66608}
  (\bibinfo {year} {2004})}\BibitemShut {NoStop}%
\bibitem [{\citenamefont {Nazarenko}(2011)}]{nazarenkobook}%
  \BibitemOpen
  \bibfield  {author} {\bibinfo {author} {\bibfnamefont {S.}~\bibnamefont
  {Nazarenko}},\ }\href@noop {} {\emph {\bibinfo {title} {Wave Turbulence}}}\
  (\bibinfo  {publisher} {Springer},\ \bibinfo {address} {Heidelberg},\
  \bibinfo {year} {2011})\BibitemShut {NoStop}%
\bibitem [{\citenamefont {Choi}\ \emph
  {et~al.}(2005{\natexlab{a}})\citenamefont {Choi}, \citenamefont {Lvov},
  \citenamefont {Nazarenko},\ and\ \citenamefont {Pokorni}}]{Choi2005361}%
  \BibitemOpen
  \bibfield  {author} {\bibinfo {author} {\bibfnamefont {Y.}~\bibnamefont
  {Choi}}, \bibinfo {author} {\bibfnamefont {Y.~V.}\ \bibnamefont {Lvov}},
  \bibinfo {author} {\bibfnamefont {S.}~\bibnamefont {Nazarenko}}, \ and\
  \bibinfo {author} {\bibfnamefont {B.}~\bibnamefont {Pokorni}},\ }\href@noop
  {} {\bibfield  {journal} {\bibinfo  {journal} {Phys. Lett. A}\ }\textbf
  {\bibinfo {volume} {339}},\ \bibinfo {pages} {361} (\bibinfo {year}
  {2005}{\natexlab{a}})}\BibitemShut {NoStop}%
\bibitem [{\citenamefont {Eyink}\ and\ \citenamefont
  {Shi}(2012)}]{Eyink20121487}%
  \BibitemOpen
  \bibfield  {author} {\bibinfo {author} {\bibfnamefont {G.~L.}\ \bibnamefont
  {Eyink}}\ and\ \bibinfo {author} {\bibfnamefont {Y.-K.}\ \bibnamefont
  {Shi}},\ }\href@noop {} {\bibfield  {journal} {\bibinfo  {journal} {Physica
  D}\ }\textbf {\bibinfo {volume} {241}},\ \bibinfo {pages} {1487} (\bibinfo
  {year} {2012})}\BibitemShut {NoStop}%
\bibitem [{\citenamefont {Benney}\ and\ \citenamefont
  {Newell}(1969)}]{benney1969random}%
  \BibitemOpen
  \bibfield  {author} {\bibinfo {author} {\bibfnamefont {D.~J.}\ \bibnamefont
  {Benney}}\ and\ \bibinfo {author} {\bibfnamefont {A.~C.}\ \bibnamefont
  {Newell}},\ }\href@noop {} {\bibfield  {journal} {\bibinfo  {journal} {Stud.
  Appl. Math}\ }\textbf {\bibinfo {volume} {48}},\ \bibinfo {pages} {29}
  (\bibinfo {year} {1969})}\BibitemShut {NoStop}%
\bibitem [{\citenamefont {Benney}\ and\ \citenamefont
  {Saffman}(1966)}]{Benney25011966}%
  \BibitemOpen
  \bibfield  {author} {\bibinfo {author} {\bibfnamefont {D.~J.}\ \bibnamefont
  {Benney}}\ and\ \bibinfo {author} {\bibfnamefont {P.~G.}\ \bibnamefont
  {Saffman}},\ }\href@noop {} {\bibfield  {journal} {\bibinfo  {journal} {Proc.
  R. Soc. A}\ }\textbf {\bibinfo {volume} {289}},\ \bibinfo {pages} {301}
  (\bibinfo {year} {1966})}\BibitemShut {NoStop}%
\bibitem [{\citenamefont {Canuto}\ \emph {et~al.}(1988)\citenamefont {Canuto},
  \citenamefont {Hussaini}, \citenamefont {Quarteroni},\ and\ \citenamefont
  {Zang}}]{Canuto}%
  \BibitemOpen
  \bibfield  {author} {\bibinfo {author} {\bibfnamefont {C.}~\bibnamefont
  {Canuto}}, \bibinfo {author} {\bibfnamefont {M.~Y.}\ \bibnamefont
  {Hussaini}}, \bibinfo {author} {\bibfnamefont {A.}~\bibnamefont
  {Quarteroni}}, \ and\ \bibinfo {author} {\bibfnamefont {T.~A.}\ \bibnamefont
  {Zang}},\ }\href@noop {} {\emph {\bibinfo {title} {Spectral Methods in Fluid
  Dynamics}}}\ (\bibinfo  {publisher} {Springer-Verlag},\ \bibinfo {year}
  {1988})\BibitemShut {NoStop}%
\bibitem [{Note1()}]{Note1}%
  \BibitemOpen
  \bibinfo {note} {The explicit expression of the solution of Eq.~(\ref
  {eqn:equation for F}) in Ref.~\cite {lvov2004nsl} is incorrect.}\BibitemShut
  {Stop}%
\bibitem [{\citenamefont {Tanaka}\ and\ \citenamefont
  {Yokoyama}(2011)}]{MT_NY_physd}%
  \BibitemOpen
  \bibfield  {author} {\bibinfo {author} {\bibfnamefont {M.}~\bibnamefont
  {Tanaka}}\ and\ \bibinfo {author} {\bibfnamefont {N.}~\bibnamefont
  {Yokoyama}},\ }\href@noop {} {\bibfield  {journal} {\bibinfo  {journal}
  {Physica D}\ }\textbf {\bibinfo {volume} {240}},\ \bibinfo {pages} {1145}
  (\bibinfo {year} {2011})}\BibitemShut {NoStop}%
\bibitem [{\citenamefont {Choi}\ \emph
  {et~al.}(2005{\natexlab{b}})\citenamefont {Choi}, \citenamefont {Lvov},\ and\
  \citenamefont {Nazarenko}}]{Choi2005121}%
  \BibitemOpen
  \bibfield  {author} {\bibinfo {author} {\bibfnamefont {Y.}~\bibnamefont
  {Choi}}, \bibinfo {author} {\bibfnamefont {Y.~V.}\ \bibnamefont {Lvov}}, \
  and\ \bibinfo {author} {\bibfnamefont {S.}~\bibnamefont {Nazarenko}},\
  }\href@noop {} {\bibfield  {journal} {\bibinfo  {journal} {Physica D}\
  }\textbf {\bibinfo {volume} {201}},\ \bibinfo {pages} {121} (\bibinfo {year}
  {2005}{\natexlab{b}})}\BibitemShut {NoStop}%
\bibitem [{\citenamefont {Janssen}(2003)}]{janssen_freak}%
  \BibitemOpen
  \bibfield  {author} {\bibinfo {author} {\bibfnamefont {P.~A. E.~M.}\
  \bibnamefont {Janssen}},\ }\href@noop {} {\bibfield  {journal} {\bibinfo
  {journal} {J. Phys. Oceanogr.}\ }\textbf {\bibinfo {volume} {33}},\ \bibinfo
  {pages} {863} (\bibinfo {year} {2003})}\BibitemShut {NoStop}%
\bibitem [{\citenamefont {Tanaka}(2007)}]{doi:10.1175/JPO3029.1}%
  \BibitemOpen
  \bibfield  {author} {\bibinfo {author} {\bibfnamefont {M.}~\bibnamefont
  {Tanaka}},\ }\href@noop {} {\bibfield  {journal} {\bibinfo  {journal} {J.
  Phys. Oceanogr.}\ }\textbf {\bibinfo {volume} {37}},\ \bibinfo {pages} {1022}
  (\bibinfo {year} {2007})}\BibitemShut {NoStop}%
\end{thebibliography}
\end{document}